\documentclass[fleqn,12pt,twoside]{article}
\usepackage{espcrc1}
\usepackage{epsfig}

\usepackage{graphicx}
\usepackage[figuresright]{rotating}

\newcommand{\AmS}{{\protect\the\textfont2
  A\kern-.1667em\lower.5ex\hbox{M}\kern-.125emS}}

\title{Electromagnetic structure and weak decay of meson K in a light-front
QCD-inspired model\thanks{Work partially supported by 
the Brazilian funding agencies FAPESP and CNPq}}
\author{Fabiano~P.~Pereira
\address{Instituto de F\'isica, 
Universidade Federal Fluminense, 
24210-900, Niter\'oi,
RJ, Brazil},
{J.~P.~B.~C.~de~Melo}
\address{Universidade Cruzeiro do Sul, 
CETEC, 08060-070, S\~{a}o Paulo, SP, Brazil}
\thanks{JPBC de Melo thanks Instituto de F\'\i sica Te\'orica, 
UNESP, for supporting facilities},
T.~Frederico
\address{Instituto Tecnol\'ogico de Aeron\'autica, 12228-900, 
S\~{a}o Jos\'e
dos Campos, SP, Brazil}, and
Lauro Tomio
\address{Instituto de F\'\i sica Te\'orica, UNESP,
01405-900, S\~{a}o Paulo, SP, Brazil}
}

\begin{document}
\maketitle
\begin{abstract}
The kaon electromagnetic (e.m.) form factor is reviewed
considering a light-front constituent quark model.
In this approach, it is discussed the relevance of the
quark-antiquark pair terms for the full covariance of the e.m. current.
It is also verified, by considering a QCD dynamical model, that
a good agreement with experimental data can be obtained for the kaon
weak decay constant once a probability of about 80\% of the valence
component is taken into account.
\end{abstract}

\section{INTRODUCTION}
The kaon, as quark-antiquark bound states, is one appropriate
system to study aspects of QCD at low and intermediate energy
regions. By using quantum field theory at the light-front the
subnuclear structure can be more easily studied
~\cite{cardarelli96,Pacheco99,Bakker01}. Within the light-front
framework and an appropriate choice of the frame, it is possible
to obtain the pion electromagnetic form factor at both space- and
time-like regimes\cite{Pacheco2006}. Using the light-cone
components $J_K^+= J^0 + J^3$ and $J_K^-= J^0 - J^3$ of the kaon
electromagnetic current, one can obtain the corresponding form
factors in the light-front formalism, with a pseudoscalar coupling
for the quarks and considering the Breit frame ($q^+=0$,
$\vec{q}_{\perp}=(q_x,0) \neq 0$)~\cite{Fabiano05}. In the case of
$J_K^+$ there is no pair term contribution in the Breit frame.
However, for the $J_K^-$ component of the electromagnetic current,
the pair term contribution is different from zero and necessary
to preserve the rotational symmetry of the current.

In the next section, we outline the main equations of the model
for the kaon electromagnetic current, detailed in
\cite{Fabiano05}, with the corresponding results obtained for the
kaon elastic form factor. In section 3, we briefly review a QCD
inspired model, presenting results for the weak decay pseudoscalar
constants compared to data. In section 4 we present our
conclusions.

\section{ELECTROMAGNETIC FORM FACTOR}

The initial light-front wave function considered in the present
model is given by: \begin{eqnarray}
\Phi^i_{Q}(x,k_\perp)=\frac{1}{(1-x)^2}
\frac{N}{(m^2_{K^+}-M^2_{0}) (m^2_{K^+}-M^2(m_{Q},m_R))} \
,\label{fupi}
\end{eqnarray}
where $N$ is a normalization constant, $Q\equiv \{ \bar q, q \}$ is
the quark or antiquark index with $m_Q$ is the corresponding quark
mass, $m^2_{K^+}$ is the kaon mass, $x=k^+/P^+$ is the momentum
fraction, and
\begin{equation}
M^2(m_Q,m_R)\equiv \frac{k^2_\perp+m^2_Q}{x}+
\frac{(P-k)^2_\perp+m^2_R}{1-x}-P^2_\perp
\label{fupi2} ,
\end{equation}
with the free quark-mass operator given by $M^2_{0}=M^2(m_{\bar
q},m_q)$. $m_R$ is a mass constant chosen to regularize the
triangle diagram. For the corresponding final wave-functions,
$\Phi^f_{\bar q}$ and $\Phi^f_{q}$, we just need to exchange
$P\leftrightarrow P^{\prime}$ in (\ref{fupi}) and (\ref{fupi2}).
The relation between the electromagnetic current $J^{\mu}$ and the
space-like kaon electromagnetic form factor $F_{K^+}(q^2)$ is
given by $\langle P^{\prime} |J^{\mu}|P^{\prime}\rangle =
(P^{\prime} + P)^{\mu} F_{K^+}(q^2)\;.\; $ In terms of the initial
$(\Phi^i_{\bar q})$ and final $(\Phi^f_{\bar q})$ light-front wave
functions, we have {\small
\begin{eqnarray}
F^{+}_{\bar q}(q^2)&=& -e_{\bar  q} \frac{N^2 g^2 N_c}{4\pi^3 P^+}
\int \frac{d^{2} k_{\perp} d x} {x}  {\cal N}^+_{\bar q}\theta(x)
\theta(1-x) \ \Phi^{*f}_{\bar q}(x,k_{\perp})
\Phi^i_{\bar q}(x,k_{\perp})  \ , \\
\ F^{+}_{q}(q^2) & = & [\ q \ \leftrightarrow \ \bar{q}  \  \mbox{in}
\ F^{+}_{\bar q}(q^2) \ ] \  \ ,
\label{form}
\end{eqnarray}}
where $N_c$ is the color number, $g$ is the coupling constant, $e_Q$ is the
charge of quark $Q$,  and
${\cal N}^+_{\bar q}=\left. ({-1}/{4}){\rm Tr}[(\rlap\slash k +m_{\bar{q}})
\gamma^5 (\rlap\slash k-\rlap\slash P^{\prime}+m_{q})
\gamma^+ (\rlap\slash k-\rlap\slash P +m_{q})
\gamma^5 ]\right|_{k_-=\bar{k}^{-}}.\;$
In the light-front approach, beside the valence contribution, we have also
the non-valence contributions to the currents.
In the case of the $J^+$ component, the non-valence component
does not contribute to the corresponding matrix
elements~\cite{Fabiano05}.
The kaon electromagnetic form factor obtained with $J^{+}$ is the sum of
two contributions from quark and antiquark currents:
\begin{equation}
F_{K^+}^{+}(q^2)=F_{q}^{+}(q^2)+F_{\bar{q}}^{+}(q^2) \;\;\;{\rm normalized\; such\; that}\;\;F^{+}_{K^+}(0)=1.\label{fplus}
\end{equation}
 In the case that we consider the $J^{-}$ component, to
obtain the kaon electromagnetic form factor, after considering the
contribution from the interval $0<k^+<P^+$ (interval I), we need
to add a second contribution, which is originated from the pair
terms, and non-zero in the interval $P^+<k^+<P^{\prime +}$
(interval II). The contribution is obtained after a Cauchy
integral in $k^-$ is performed in the limit $P^{\prime +}\to
P^+$~\cite{Fabiano05}. So, instead of (\ref{fplus}), we will have:
\begin{equation}
F_{K^+}^{-}(q^2)= \left[
F_{q}^{-}(q^2)+F_{\bar{q}}^{-}(q^2)
\right]_{(I)}+ \left[F^{-}(q^2)\right]_{(II)},
\end{equation}
normalized by the charge conservation to $F^{-}_{K^+}(0)=1.$

The parameters of the model are the constituent quark masses,
$m_q=m_u=m_d=$ 220 MeV, $m_s=$ 419 MeV and the regulator mass
$m_R=$946 MeV, adjusted to fit the electromagnetic radius of the
kaon. The electromagnetic radius is related to the corresponding
form factor, with the mean-square-radius given by
\begin{equation}
\langle r^2_{K^+}\rangle =
6\left[\frac{dF_{K^+}(q^2)}{dq^2}\right]_{q^2=0}.
\end{equation}
With the parameters adjusted as given above, we have $\langle
r^2_{K^+}\rangle =$ 0.354 fm$^2$, which is very close to the
experimental value $\langle r^2_{K^+}\rangle|_{exp} =$ 0.340
fm$^2$~\cite{kaexp}.

Our results for the kaon electromagnetic form factor are presented 
in Fig.~1, in comparison with available experimental data~\cite{kaexp}.
We observe that the full kaon electromagnetic form factor is
covariant only after the inclusion of the pair terms or
non-valence contribution to the $J^{-}_{K^{+}}$ component of the
electromagnetic current.
\begin{figure}[thbp]
\vspace{-.7cm}
\begin{center}
\includegraphics[width=11.cm
]{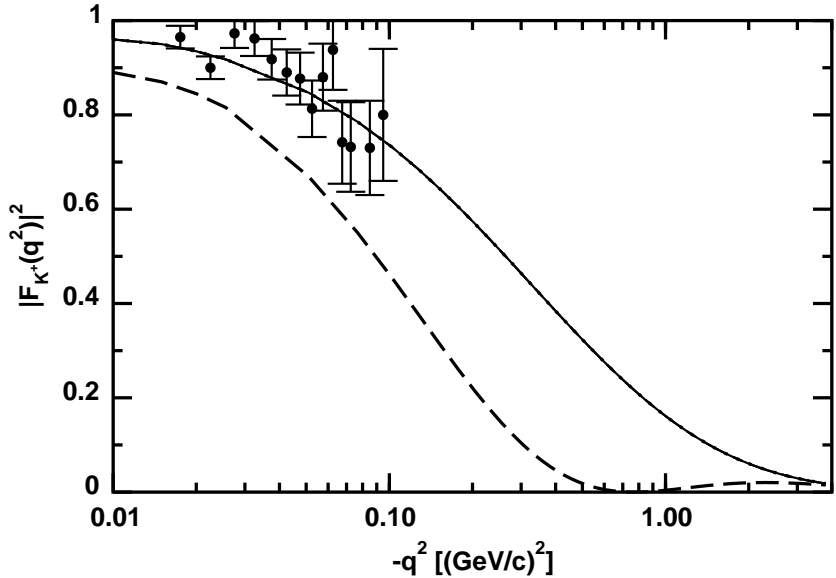}
\end{center}
 \vspace{-1.2cm}
\caption[dummy0]{ The kaon electromagnetic form factor is obtained
with the plus and minus components of the e.m. current (both cases are 
shown by the solid-line results) and compared with experimental data~\cite{kaexp}. 
The dashed-line curve 
shows the form factor without the pair terms contribution in $ J^{-}_{K^+}$.
} \label{fig1}
\vspace{-0.5cm}
\end{figure}

\section{WEAK DECAY CONSTANTS IN A QCD INSPIRED MODEL}

Next, we briefly review the calculation of the pseudoscalar constants,
in a light-front QCD-inspired dynamical model.
In this case, the constituent quark masses need to be readjusted
in view of the fact that, differently from the approach outlined in
section 2, the wave-function is obtained from an eigenvalue equation,
as follows.

The valence wave function is obtained by solving an eigenvalue
equation for the effective square mass operator
$M^2_{ps}$~\cite{Tobias2001}: {\small
\begin{eqnarray} M^2_{ps}\;\psi (x,{\vec k_\perp}) &=&
M^2_0(x,k_\perp)\;\psi (x,{\vec k_\perp})
-\int\frac{dx' d{\vec k'_\perp}\theta(x')\theta (1-x')}
{\sqrt{x(1-x)x'(1-x')}} \times
\nonumber \\ & \times &
\left(\frac{4m_1m_2}{3\pi^2}\frac{\alpha}{Q^2}-\lambda_{ps}
g(M^2_0(x,k_\perp))g(M^2_0(x',k'_\perp))\right) \psi (x',{\vec k'_\perp}) ~,
\label{p1}
\end{eqnarray}}
where $M^2_0(x,k_\perp)\equiv ({\vec k_\perp}^2+m^2_1)/{x}+({\vec
k_\perp}^2+m^2_2)/(1-x)  $
 is the free square mass operator in the meson rest frame,
 $m_{1,2}$ are the constituent quark masses, $\alpha$ gives the strength
 of the Coulomb-like interaction.
$g(K)$ is the model form factor, with $\lambda_{ps}$ the strength of the separable
interaction. We consider two expressions for the form factors:
\begin{eqnarray}
g^{(a)}(K^2)={1\over \beta^{(a)}+K^2}~~\textrm{and}~~
g^{(b)}(K^2)={1\over K^2}+\left({\beta^{(b)}\over K^2}\right)^2 ,
\label{ff}
\end{eqnarray}
where the parameters $\beta^{(a,b)}$ and $\lambda_{ps}$ are
adjusted to reproduce the experimental values of the pion
electromagnetic radius and  mass, $m_{\pi}$.  For $\alpha=0.5$, we
have $\beta^{(a)}=-(634.5~{\rm MeV})^2$ and
$\beta^{(b)}=(1171~{\rm MeV})^2$. $m_u=384~{\rm MeV},~m_s=508~{\rm
MeV}$. In Table~1, we have the results compared with experimental
data~\cite{pdg}.
\begin{table}[t,b,h]
\caption
{Results for the kaon and pion weak decay constants,
compared with experimental data.
The model is adjusted to reproduce pion radius and mass.}
\begin{center}
\begin{tabular}{c|ccc|cc}
\hline\hline $q \overline q$ &$f^{(a)}_{ps}$(MeV)& $f^{(b)}_{ps}$(MeV)&
$f^{exp}_{ps}$(MeV)&
$M^{(a)}_{ps}$(MeV)&
$M^{exp}_{ps}$(MeV)~\cite{pdg}
\\ \hline \hline
$\pi^+(u\overline d)$ & 110 & 110 &$92.4\pm.07\pm 0.25$
\cite{pdg} & 140 & 140\\
\hline $K^+(u\overline s) $  & 126 & 121&
$113.0\pm1.0\pm0.31$\cite{pdg} & 490 & 494  \\
\hline\hline
\end{tabular} \\
\vspace{-.5cm}
\end{center}\label{table1}
\vspace{-0.5cm}
\end{table}

\section{CONCLUSIONS}
Considering a light-front model wave-function we have observed a
good agreement of the results for the kaon electromagnetic form
factor with experimental data. The electromagnetic form factor was
obtained using the plus and minus components of the
electromagnetic current. The inclusion of the non-valence
component of the current was shown to be essential in this
approach to obtain covariant results for the calculated matrix
elements. We also show that a good agreement with experimental
data is obtained for the kaon weak decay constants once a
probability of the valence component of about 80\% is taken into
account.


\begin{thebibliography}{99}
\bibitem{cardarelli96} F. Cardarelli, I. L. Grach, I.~M.~Narodetsky,
E.~Pace, G.~Salme, S.~Simula, Phys.~Rev.~D 53 (1996) 6682.

\bibitem{Pacheco99} J. P. B. C. de Melo,
H. W. Naus and T. Frederico,
Phy.~Rev.~C 59~(1999)~2278.

\bibitem{Bakker01} B. L. G. Bakker, H.-M. Choi and
C.-R. Ji, Phys. Rev. D 63 (2001) 074014.

\bibitem{Pacheco2006}
J. P. B. C. de Melo,
T. Frederico, E. Pace and G. Salm\`e,
Phy. Rev. D 73 (2006) 074013;
J. P. B. C. de Melo, T. Frederico, E. Pace and G. Salm\`e,
Phy. Lett. B 581 (2004) 75.

\bibitem{Fabiano05} F.P.~Pereira, J.P.B.C.~de Melo,
T.~Frederico and L.~Tomio, Phys.~of Part. and Nucl.
36 (2005) 5217; F.P.~Pereira,
{\it Fatores de Forma Eletromagn\'eticos do P\'\i on
e do Kaon na Frente de Luz}, Msc Dissertation, IFT, S\~ao Paulo, 2005.

\bibitem{kaexp}
S. R. Amendolia et al., Phys. Lett. B 178 (1986) 435.

\bibitem{Tobias2001} T. Frederico and H.-C. Pauli,
Phy. Rev. D 64 (2001) 054004; L. A. M. Salcedo, J. P. B. C. de
Melo, D. Hadjmichef and T. Frederico, Eur. Phys. J. A 27 (2006)
213.

\bibitem{pdg}
W.-M. Yao et al., Journal of Physics G 33 (2006) 1.

\end{thebibliography}
\end{document}